\documentstyle[prl,aps,twocolumn,psfig]{revtex}

\def\Ai{{\rm Ai}}
\def\Bi{{\rm Bi}}
\begin{document}

\twocolumn[\hsize\textwidth\columnwidth\hsize\csname
@twocolumnfalse\endcsname


\title{Selected elevation in quantum tunneling}
\author{Er'el Granot \footnotemark}

\address{School of Physics and Astronomy, Raymond and Beverly Sackler Faculty of Exact Sciences, Tel-Aviv University, 69978 Tel Aviv, Israel}

\date{\today}
\maketitle
\begin{abstract}
\begin{quote}
\parbox{16 cm}{\small
The tunneling through an opaque barrier with a {\em strong}
oscillating component is investigated. It is shown, that in the
strong perturbations regime (in contrast to the weak one), higher
perturbations rate does not necessarily improve the activation. In
fact, in this regime two rival factors play a role, and as a
consequence, this tunneling system behaves like a sensitive
frequency-shifter device: for most incident particles' energies
activation occurs and the particles are {\em energetically
elevated }, while for specific energies activation is depressed
and the transmission is very low. This effect is unique to the
{\em strong} perturbation regime, and it is {\em totally absent}
in the weak perturbation case. Moreover, it cannot be deduced even
in the adiabatic regime. It is conjectured that this mechanism can
be used as a frequency-dependent transistor, in which the device's
transmission is governed by the external field {\em frequency}.}
\end{quote}
\end{abstract}

\pacs{PACS: 73.40G and 66.35}

]

\narrowtext \footnotetext{erel.g@kailight.com} \noindent When a
quantum particle propagates through an opaque barrier in the
classically forbidden domain, it tunnels. This conduct is
manifested by the exponentially small transmission probability.
Nevertheless, when some part of the barrier weakly oscillates the
particle will be activated to higher energies, and the
transmission will increase substantially
\cite{Buttiker_82}-\cite{Fisher}. However, when the temporal
change is not merely a perturbation but rather a strong variation
the dynamics are too complicated to be described by such a simple
rule of thumb, and most of the interesting phenomena belong to
this category.

While it is well known that even a very weak external oscillating
field may considerably increase the tunneling current
\cite{Ivlev,Fisher}, changes which are comparable to the initial
system parameters can cause elevator resonant activation and
activation assisted tunneling \cite{Azbel_92_91},\cite{Azbel_92},
charge quantization pumping \cite{Aleiner}, as well as wave
function collapse \cite{Azbel_94}. That is, large changes reveal a
wealth of physical phenomena.

When the tunneling particle energy ($\Omega $) is close to the
potential barrier height ($V$), the dynamics become more
complicated since the perturbation amplitude ($\Delta V$) can
exceed $V-\Omega $ , i.e., the perturbation ''strength'' may be
larger than the effective tunneling barrier. This case is
extremely sensitive, since the dynamics are governed by two rival
factors. On the one hand, the oscillation's amplitude is so large,
that for a finite segment of the oscillation period the
alternating potential blocks the particles' transmission. Hence,
following this reasoning, the tunneling rate should decrease. On
the other hand, energy quanta generated by the oscillating barrier
can be absorbed by the tunneling particle to assist it in the due
course of tunneling.

In this work, we investigate the interplay between these two rival
factors, and show that the competition between them is responsible
for the high sensitivity to the system's parameters. For example,
it so happens that the incident particles are not always activated
to higher energy: when their initial energy is equal to one of a
series of specific energies the activation is frustrated,
elevation to higher energies does not occur and the transmission
is decreased accordingly.

The problem of tunneling in the presence of an external
time-dependent potential has been extensively studied
\cite{Buttiker_82}-\cite{Azbel_94}. However, in this paper the
strong perturbation regime is addressed and no approximate
assumptions are made in the solution analysis. We give this case
both an {\em exact numerical} solution, and an {\em analytical}
solution, using an approximation, to show that the general
conclusions can be deduced even in the slowly varying regime.
However, we also show that the oscillations rate cannot be
arbitrarily small, and in fact it should be larger than the
spectral bandwidth of the resonances, otherwise (and, in
particular, the adiabatic regime) the effect vanishes.

The apparent underlying physics can be described as follows: Due
to the oscillations, the alternating barrier functions as an
impenetrable barrier (or at least a very opaque one) only for a
fraction of the oscillating period and therefore, a time-window
(denoted $\tau _{W}$) is formed in which the incident particle can
propagate. Consequently, in order to prevent a destructive
pattern, the incoming particles' frequency ,$\Omega $ (in units
where $\hbar=1$) should be equal to an integer of both the oscillations' frequency $\omega $%
\ and $2\pi /\tau _{W}$ (i.e., $\Omega =2\pi m_{1}/\tau _{W}$ and
$\Omega =m_{2}\omega $\ for integer $m_{1}$ and $m_{2}$).
Therefore, a selection rule for activation should be expected .
However, this selection rule is rather complicated since $\tau
_{W}$ is a function of the incident particles' frequency ($\Omega
$). It turns out that the energies, for which activation occurs,
have a nontrivial dependence on the integer $m_{1}$ (roughly like
$m_{1}^{2/3}$).

A few words should be added here about the plausibility of the
effect's practical implementation. The industry has a special
interest in resonant tunneling devices(RTD), which allow
miniaturizing electronic circuits and improving their performance.
Ordinary RTD are very sensitive to manufacturing processes,
temperature, and impurities, and as a consequence a reliable
device is highly costly. The presented tunneling device allows for
the fabrication of a low-cost device where any resonant refinement
can be done by variations in the external field \emph{frequency},
and no special geometry or manufacturing restrictions are needed.

In this paper we discuss the tunneling dynamics of a beam of
particles which are activated by strong harmonic perturbations.
The tunneling takes place through a very opaque potential barrier
(high and wide). In order to have the barrier opaque at all times,
we follow \cite{Azbel_92_91} and discuss the extreme case of an
oscillating point potential $-\beta \delta \left( x\right) \cos
\left( \omega t+\eta \right) $ (see Fig.1).

\begin{figure}
\psfig{figure=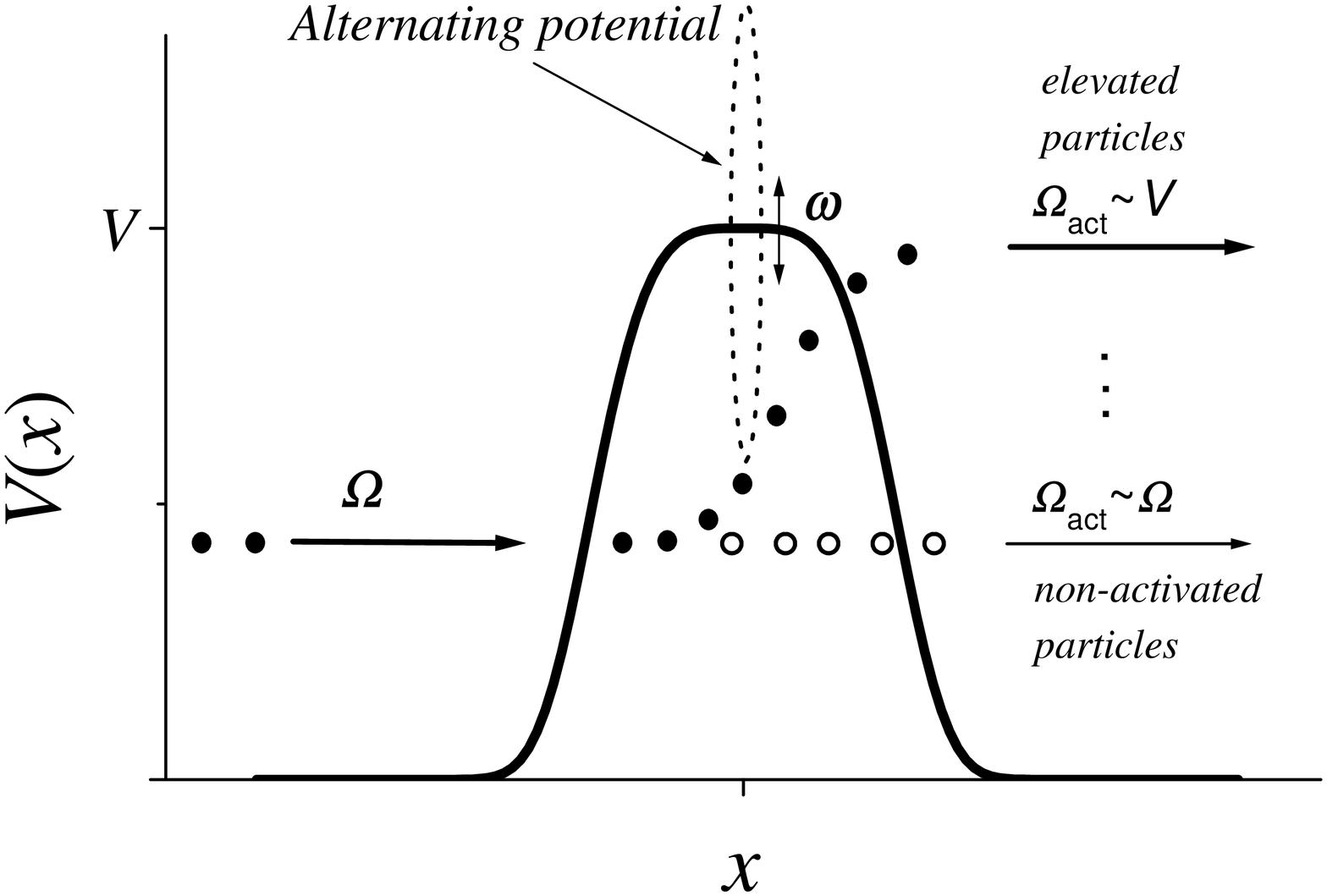,width=8cm,bbllx=20bp,bblly=10bp,bburx=780bp,bbury=550bp,clip=}
\caption{\small \it An illustration of the system: The incidnet
particles' energy $\Omega$ (and the oscillating frequency
$\omega$) determines whether the incident particles will be
elevated to higher energy states (dark circles - most cases) or
will remain in their initial state (open circles - specific
cases).} \label{figure1}
\end{figure}

In terms of the Schr\"{o}dinger equation, the dynamics can be
expressed by

\begin{equation}
-\psi ^{\prime \prime }-\beta \delta \left( x\right) \cos \left( \omega
t+\eta \right) \psi +V\left( x\right) \psi =i\dot{\psi}  \label{1}
\end{equation}

where we adopt the units $2m=1$,$\hbar =1$, and the notations $\psi ^{\prime
\prime }\equiv \partial ^{2}\psi /\partial x^{2}$, $\dot{\psi}\equiv
\partial \psi /\partial t$ , and $V\left( x\right) $\ is the potential
barrier which vanishes quickly for $\left| x\right| \rightarrow \infty $.

The time-dependent solution can be written as a discrete Fourier
transform

\begin{equation}
\psi \left( x<0\right) =\varphi _{\Omega }^{+}e^{-i\Omega
t}+\sum_{n}r_{n}\varphi _{\Omega +n\omega }^{-}e^{-i\left( \Omega
+n\omega \right) t}  \label{2a}
\end{equation}

\begin{equation}
\psi \left( x>0\right) =\sum_{n}t_{n}\varphi _{\Omega +n\omega
}^{+}e^{-i\left( \Omega +n\omega \right) t}  \label{2b}
\end{equation}

where $\varphi_{\omega }^{\pm }$ are the solutions of the
stationary state Schr\"{o}dinger equation, which does not include
the oscillating term, i.e.

\begin{equation}
-\varphi _{\omega }^{\pm \prime \prime }+\left[ V(x)-\omega\right]
\varphi _{\omega }^{\pm }=0
\label{3}
\end{equation}

The homogeneous solutions $\varphi _{\omega }^{+}$\ describe waves
that propagate from $-\infty $ to $+\infty $, while $\varphi
_{\omega }^{-}$
describes the waves that are incoming from $+\infty $ and outgoing to $%
-\infty $. Thus, $\varphi _{\omega }^{+}\rightarrow \tau _{\omega }e^{i\sqrt{%
\omega }x-i\omega t}$ (for $x\rightarrow \infty $) while $\varphi _{\omega
}^{-}\rightarrow \tau _{\omega }e^{-i\sqrt{\omega }x-i\omega t}$ (for $%
x\rightarrow -\infty $) , and $\left| \tau _{\omega }\right|
^{2}$\ is the probability of penetrating the barrier.

By taking care of the matching conditions of the solutions in eqs.
\ref{2a} and \ref{2b} at $x=0$, we easily obtain

\begin{equation}
2s_{n}\chi _{n}+\beta \left( s_{n-1}+s_{n+1}\right) =2\chi
_{0}\varphi _{0}^{+}\delta _{n0}.  \label{6}
\end{equation}

When using the following notations

\begin{equation}
\chi _{n}\equiv \frac{\varphi _{n}^{+\prime }}{\varphi _{n}^{+}}-\frac{%
\varphi _{n}^{-\prime }}{\varphi _{n}^{-}}
\end{equation}

and

\begin{equation}
s_{n}\equiv e^{i\eta n}\varphi _{n}^{+}t_{n},  \label{5}
\end{equation}

where $\varphi _{n}^{\pm }\equiv \varphi _{\Omega +n\omega }^{\pm
}\left( x=0\right) $, $\varphi _{n}^{\pm \prime }\equiv \partial
\varphi _{\Omega +n\omega }^{\pm }/\partial x|_{x=0}$, $\delta
_{n0}$ is the Kronecker delta and $t_{n}\equiv t_{\Omega +n\omega
}$.

This difference equation can readily be solved. Notice that
\begin{equation}
\chi _{n}=1/g_{n}\left( 0\right)  \label{7a}
\end{equation}

where $g_{n}\left( x\right) $ is the Green function of the equation $-\psi
^{\prime \prime }+V\left( x\right) \psi =\left( \Omega +n\omega \right) \psi
$.

Thus, in the case of a perfectly symmetric rectangular barrier,
$\chi _{n}$ comes directly from \cite{Merzbacher_70}

\begin{equation}
g_{n}\left( 0\right) =-\coth \left[ \rho _{n}L+i\arctan \left(
k_{n}/\rho _{n}\right) \right]/2\rho _{n}  \label{7b}
\end{equation}

where $k_{n}\equiv \sqrt{\Omega +n\omega }$, $\rho _{n}\equiv \sqrt{%
V-k_{n}^{2}}$, $2L$ is the barrier length and finally $V$ is its
potential height.

In Fig.2 we present the exact numerical solution of eq.\ref{6} in
the case of a rectangular barrier (in the figure the absolute
value of $s_{n}$ is presented), which is the spectral solution of
eq.\ref{1} at $x=0$. This figures illustrates the solution's
sensitivity to the incoming particles' energy: a 2\% change in
$\Omega $ causes a severe reduction in activation.

To formulate an analytical expression for this solution, we take
advantage of the fact that the perturbations are strong, i.e., we
can assume that $\beta ^{2}\gg V-\Omega \gg \omega $. Moreover,
although in the numerical analysis we used the exact form of the
Green function (eq.\ref{7b}), in the case of an opaque barrier,
the approximation $g_{n}\left( 0\right) \simeq -\left( 2\rho
_{n}\right) ^{-1}$ may be used with great accuracy.

In this strong perturbation amplitude and low-frequency regime,
the difference equation (eq.\ref{6}) can be transformed to a
differential equation. By using the definitions
\begin{equation}
G\left( n\right) \equiv \beta s_{n}/\chi _{0}  \label{8}
\end{equation}

and $ c\left( n\right) \equiv 1+\chi _{n}/\beta , $ one easily
obtains the differential equation

\begin{equation}
\frac{d^{2}}{dn^{2}}G\left( n\right) +c\left( n\right) G\left( n\right)
=\delta \left( n\right)  \label{9}
\end{equation}

where $\delta \left( n\right) $ is the Dirac delta function.

Hence, one can regard $G\left( n\right) $ as having a Green function
properties (including the singularity at $n=0$).

\begin{figure}
\psfig{figure=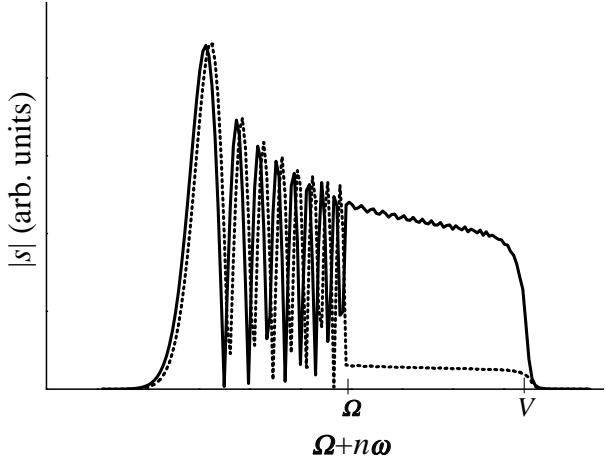,width=8cm,bbllx=85bp,bblly=10bp,bburx=700bp,bbury=500bp,clip=}
\caption{\small \it A plot of the transmission coefficient $\left|
s_{n}\right| $ (defined in Eq.\ref{5}) as a function of the
transmitted particles frequency (the activated energy) $\Omega
+n\omega $. The solid line represents the case $\Omega /V=0.625$
and the dotted one represents the case $\Omega /V=0.6125$. The
other parameters are $\omega /V=0.0075$, $\sqrt{V}L=10.75$\ and
$\beta L=9.35$.} \label{figure2}
\end{figure}

When the Green function $G\left( n\right) $ is known the
transmitted
solution of eq.\ref{1} follows directly from eqs.\ref{2b},\ref{5} and \ref{8}%
:

\begin{equation}
\psi \left( x\geq 0\right) =\frac{\chi _{0}\varphi _{0}^{+}}{\beta }%
e^{-i\Omega t}\int dnG\left( n\right) \frac{\varphi _{\Omega +n\omega
}^{+}(x)}{\varphi _{\Omega +n\omega }^{+}(0)}e^{-in\left( \omega t+\eta
\right) }  \label{10}
\end{equation}

In particular, the scattered wave function at $x=0$ is
proportional to the Fourier transform of the Green function.

Since $V-\Omega \gg \omega $, one can approximate

\begin{equation}
c\simeq 1-2\rho /\beta +\omega n/\left( \beta \rho \right) \text{,}
\label{12}
\end{equation}

where $\rho \equiv \sqrt{V-\Omega }$.

Then, we can define for convenience the variable
\begin{equation}
\xi \equiv \left( n+\rho \frac{\beta -2\rho }{\omega }\right) \left( \frac{%
\omega }{2\beta \rho }\right) ^{1/3}  \label{13}
\end{equation}

and the Green function is then

\begin{equation}
\begin{array}{rl}
 G\left( \xi \right) =&
 -i\pi\left( \frac{2\beta \rho }{\omega
}\right) ^{1/3}\times \\
&\left\{
\begin{array}{lc}
\Ai\left( -\xi \right) \left[ \Ai\left( -\xi _{0}\right)
+i\Bi\left( -\xi
_{0}\right) \right]  & \text{for \ }\xi <\xi _{0} \\
\Ai\left( -\xi _{0}\right) \left[ \Ai\left( -\xi \right)
+i\Bi\left( -\xi \right) \right]  & \text{for \ }\xi >\xi _{0}
\end{array}
\right.
\end{array}
  \label{14}
\end{equation}

where $\xi _{0}\equiv \xi \left( n=0\right) =\left( 1-2\rho /\beta
\right) \left( \beta \rho /\omega \sqrt{2} \right) ^{2/3}$ and
$\Ai$ and $\Bi$ are the Airy functions of the first and second
kind, respectively.

For frequencies which are lower than $\Omega $ (negative $n$' s) the {\em %
amplitude} oscillates like a simple Airy function (see Fig.2 for a
typical plot of $\left| s_{n}\right| $ , which is related to the
Green function by eq.\ref{8})

\bigskip
\begin{equation}
\left| G\left( \xi \right) \right| ^{2}\simeq \pi \left( 2\beta
\rho /\omega \right) ^{2/3}\xi _{0}^{-1/2}\Ai^{2}\left( -\xi
\right) \text{ \ \ \ \ \ \ \ \ for \ \ }n <0
\end{equation}

since $\left| \Ai\left( -\xi _{0}\right) +i\Bi\left( -\xi
_{0}\right) \right| ^{2}\simeq \pi ^{-1}\xi _{0}^{-1/2}$. That
explains the insensitivity of the amplitude of $G$ (for a specific
$n<0$ or $\xi <\xi _{0} $ ) to small variations in the incoming
energy $\Omega $ (see, for example,
Fig.2). However, for a specific incoming energy $\Omega $, the amplitude of $%
G$ oscillates with respect to the {\em transmitted} energies (i.e., $\Omega
+n\omega $). In this regime (i.e., $\xi >\xi _{0}$)
\begin{equation}
\left| G\left( \xi \right) \right| ^{2}\simeq \pi \left( 2\beta
\rho /\omega \right) ^{2/3}\xi ^{-1/2}\Ai^{2}\left( -\xi
_{0}\right) \text{ \ \ \ \ \ \ for \ \ }n>0  \label{16}
\end{equation}
\qquad

which means that for an incident particles' energy $\Omega $, the
amplitude of the Green function has a very mild dependence on the
{\em transmitted} particles' energies (i.e., on $n$), while it is
strongly dependent on the {\em incident} particles' energies
$\Omega $. This can explain the result presented in Fig.2, where a
two percent change in the incoming particles' energy was enough to
frustrate the activation to higher energies.

It is clear from eq.\ref{16} and from the periodicity of the Airy
function, that the probability of an incident particle being
activated to higher energy is very sensitive to its initial
energy. This sensitivity is manifested in the following
calculation of the mean activation energy:

\begin{equation}
\Omega _{act}=\Omega +\omega \left| \frac{\chi _{0}\varphi _{0}^{+}}{\beta }%
\right| \int ndn\left| G\left( n\right) \frac{\varphi _{\Omega +n\omega
}^{+}\left( x>L\right) }{\varphi _{\Omega +n\omega }^{+}\left( 0\right) }%
\right| ^{2}  \label{20}
\end{equation}

A plot of the mean activation energy ($\Omega _{act}$) as a
function of the incident one ($\Omega $) is shown in Fig.3 (an
exact numeric solution).

The opaqueness of the barrier is the cause for the sharp changes in $\Omega
_{act}$. Since the tunneling coefficient $\varphi _{\Omega +n\omega
}^{+}\left( x>L\right) /\varphi _{\Omega +n\omega }^{+}\left( 0\right)
\simeq \exp \left( -\sqrt{V-\Omega -n\omega }L\right) $ is exponentially
small for non-activated particles, there is a great advantage for a particle
to be activated to higher energies $\Omega +n\omega \simeq V$. However,
since $G\left( n\right) $ is an oscillating function of the incoming energy $%
\Omega $, there are some energies $\Omega _{m}$ for which $G\left(
\left( V-\Omega _{m}\right) /\omega \right) $ vanishes. In these
cases, according to eq. \ref{16}, not only is the transition to
energy $V$\ forbidden, but the transition to all the other higher energies (which correspond to $%
n>0$) is, as well.

Therefore, the particles must tunnel out with energy which is very
close to their initial one (i.e., $\Omega_{act}=\Omega$). That
explains the source of these oscillations and the sharp
transitions between the activated energies $\simeq V$ and the
non-activated ones $\simeq \Omega $. The transitions occur when
the Airy function of eq.\ref{16} vanishes or, more accurately,
when $\Ai^{2}\left( -\xi _{0}\right) \simeq \exp \left(
-\sqrt{V-\Omega _{m}}L\right) $. Taking the low-frequency limit
($\omega \rightarrow 0$), the Airy function can be expressed by
simple trigonometric functions to obtain the transition criterion
$\cos ^{2}\left[ \frac{2}{3}\left( 1-\frac{2\rho _{m}}{\beta }%
\right) ^{3/2}\frac{\rho \beta }{\omega }-\frac{\pi }{4}\right]
\simeq \exp \left( -\rho _{m}L\right) $ where $\rho _{m}\equiv
\sqrt{V-\Omega _{m}}$. Therefore, the incoming energies for which
$\Omega _{act}\simeq \Omega $, and thus no activation occurs, are
approximately

\begin{equation}
\Omega _{m}\simeq V-\beta ^{2}/4+\frac{1}{2}\left[
\frac{3}{2}\beta \omega \left( m+\frac{3}{4}\right) \pi \right]
^{2/3}  \label{21}
\end{equation}

and the spectral width of these regions is exponentially small
$\Gamma \simeq \exp \left( -\beta L/2\right) $. In Fig.3
the non-activated energies $\Omega _{m}$, for which $\Omega _{act}\simeq \Omega $%
, are the minima of the plot while the maxima correspond to full
activation (i.e., $\Omega _{act}\simeq V$).

\begin{figure}
\psfig{figure=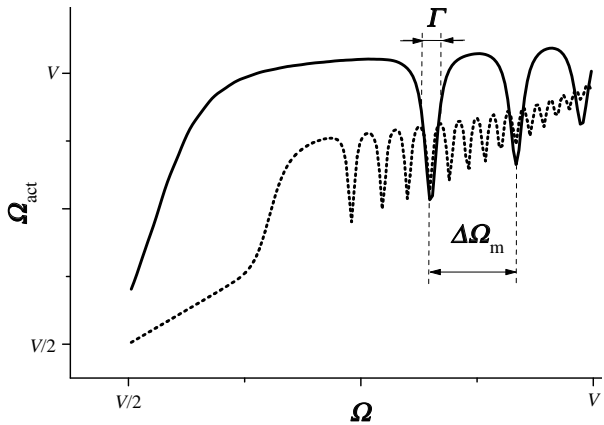,width=8cm,bbllx=55bp,bblly=1bp,bburx=700bp,bbury=500bp,clip=}
\caption{\small \it Two characteristics plots (for arbitrary
parameters) of the mean activation energy ($\Omega _{act}$) as a
function of the incident particles' energy $\Omega $\ (see
Eq.\ref{20}). The two curves represents the same system except for
$\omega$, which is five times smaller in the dotted curve}
\label{figure3}
\end{figure}

It should be emphasized here that $\Delta_{m}$ are always much
larger than the oscillations' frequency $\omega$, and therefore
the resonances pattern are not related to the oscillator quanta
(see, for example, ref. \cite{Bagwell_92}).

This effect occurs {\em only} when the perturbations are strong;
otherwise (i.e., $\beta ^{2}\ll V-\Omega $), the solution of eq.
\ref{6} is reduced to $s_{n}\sim \prod\nolimits_{n^{\prime
}=1}^{n}\beta /2\chi _{n^{\prime }}$, which is always (for large
$n$) an exponentially small quantity. In Fig.3 one can see that
when the requirement $\beta ^{2}\gg V-\Omega $ does not hold the
effect vanishes.

Moreover, when the oscillations rate decreases (i.e., $\omega$
decreases) the spectral difference between two successive valleys
shrinkage, i.e., $\Delta\Omega_{m} \equiv\Omega_{m}-\Omega_{m-1}$
also decreases. It is therefore clear, that when $\omega$ is small
enough so that $\Delta\Omega_{m}<\Gamma$ the effect vanishes and
elevation to higher energies is depressed. In Fig.3 this is shown
in the dotted line.

Hence, this effect cannot be anticipated neither in the adiabatic
regime nor in the weak perturbation approach. In both regimes the
effect vanishes.

In general, the exact geometry and shape of the barrier and the
impurity are of no essential importance (see, e.g., generalization
in ref.\cite{Azbel_92}). One can always (for any given geometry)
control the activation energy and the transmission by adapting the
external field frequency.

\bigskip

To summarize, tunneling transmission through an opaque barrier
with an oscillating section was investigated. It was shown that in
the \emph{strong} perturbation (and \emph{non} adiabatic) regime a
new selection-rule appears. Not all the incident particles are
activated, as could be anticipated in a tunneling process.
Although in most cases the incident particles are elevated to
higher energy states, for some incident particles' energy
activation is depressed and the particles remain approximately in
their initial states. The spectral width of these energy domains
is exponentially small.

I would like to thank Prof. Mark Azbel for enlightening
discussions.

\end{document}